\documentstyle[aps,preprint,epsf,floats,eqsecnum]{revtex}

\def\bfk{{\bbox k}} 
\def\bfp{{\bbox p}}

\newcommand{\lsim}{\raisebox{-0.7ex}{$\stackrel{\textstyle <}{\sim}$ }}
\def\darr#1{\raise1.5ex\hbox{$\leftrightarrow$}\mkern-16.5mu #1}
\def\){\right)} 
\def\({\left(} 
\def\]{\right]} 
\def\[{\left[}

\def\DL{\stackrel{\leftarrow}{\bf D}}
\def\DR{\stackrel{\rightarrow}{\bf D}}
\def\CC{{\tilde C}}
\def\Align{&=&}

\def\MVD{\frac{M_N}{4\pi}\frac{\gamma^2}{\frac{1}{a^{exp}}+ip}}
\def\MVr{\frac{r_0}{2}\frac{p^2}{\frac{1}{a^{exp}}+ip}}
\def\t{{({}^3S_1)}}

\newcommand{\eqn}[1]{\label{eq:#1}}
\newcommand{\refeq}[1]{(\ref{eq:#1})}

\newcommand{\beq}{\begin{eqnarray}}
\newcommand{\eeq}{\end{eqnarray}}

\newcommand{\mcal}[1]{{\mathcal #1}}
\newcommand{\makefig}[4]{\begin{figure}[t] 
                           \centerline{\epsfysize=#3 in \epsfbox{#2}} 
                           \caption{#4 \label{#1}} 
                         \end{figure}}

\def\Journal#1#2#3#4{{#1} {\bf #2}, #3 (#4)}

\def\NPB{{\em Nucl. Phys.} B}
\def\NPA{{\em Nucl. Phys.} A}
\def\NP{{\em Nucl. Phys.} }
\def\PLB{{\em Phys. Lett.} B}
\def\PRL{\em Phys. Rev. Lett.}

\def\PRC{{\em Phys. Rev.} C}

\def\PR{{\em Phys. Rev.} }
\def\ZPA{{\em Z. Phys.} A}

\def\AJSS{\em Astrophys. J. Suppl. Ser.}
\def\ARAnA{\em Annual Rev. of Astronomy and Astrophys.}

\tighten

\begin{document}

\preprint{NT@UW-99-54}

\title{ Precision Calculation of $np\rightarrow d\gamma$ Cross Section for
 Big-Bang Nucleosynthesis}
\author{Gautam Rupak} 
\address{ Department of Physics,\\ University of Washington,\\ Seattle, 
WA 98195}

\maketitle

\begin{abstract}
{An effective field theory calculation of the $np\rightarrow d\gamma$ cross 
section accurate to
 $1\%$ for center of mass energies $E\lsim 1$ MeV is presented. At these 
energies, which are relevant for big-bang nucleosynthesis, isovector magnetic 
transitions $M1_V$ and isovector electric transitions $E1_V$ give the 
dominant contributions. The $M1_V$ amplitude is calculated up to 
next-to-next-to-leading order (N$^2$LO), and the contribution from 
the associated 
four-nucleon-one-photon operator is determined from the cold neutron 
capture 
rate. The $E1_V$ amplitude is calculated up to N$^4$LO. The 
four-nucleon-one-photon operator contribution to $E1_V$ is determined from 
the related deuteron photodisintegration reaction $\gamma d\rightarrow np$.}
\end{abstract}

\vskip 3in

\leftline{September 1999}
%
%
%
%
\vfill\eject

\begin{section}{Introduction}
\label{intro}
Big-bang nucleosynthesis (BBN) is a cornerstone of big-bang 
cosmology~\cite{BNTT}. Primordial deuterium production is very sensitive to 
the baryon density of the universe and thus the BBN prediction of deuterium
abundance can be used to infer this
baryon density. These deuterium abundance calculations use the cross section 
for $np\rightarrow d\gamma$ as one of the inputs. Thus, an accurate estimation
of the  
$np\rightarrow d\gamma$ cross section is essential to the BBN prediction of 
deuterium and other light element abundances.  At the energies  
relevant for BBN, $0.02\lsim E\lsim 0.2$ MeV, 
this reaction is not well-measured experimentally and there 
are significant theoretical uncertainties~\cite{BNTT}. For example, in the 
model calculation of Smith, Kawano and Malaney (SKM)  
an error of $5\%$ 
was assigned to the cross section for $np \rightarrow d\gamma$~\cite{SKM}. 
The SKM result agrees with a slightly earlier evaluation 
from ENDF/B-VI~\cite{ENDF}.
 The errors in this calculation are not well
documented and they could be as large as $10$-$15\%$~\cite{hale}. 
There is a much
earlier calculation (1967) by Fowler, Caughlan and Zimmerman~\cite{FCZ}
 which agrees with the SKM and
ENDF/B-VI result for energies $E\lsim 0.1$ MeV but disagrees significantly  
at higher energies~\cite{SKM}.   
A recent model
 independent calculation by Chen and Savage, using a low energy 
effective field theory (EFT) predicted a theoretical uncertainty of
$4\%$~\cite{CS}. 
This work is a one higher order calculation in the perturbative expansion of 
the EFT used in Ref.~\cite{CS}. The theoretical uncertainty is estimated to be
 $\lsim 1\%$ for center of mass energies $E\lsim 1$ MeV.

 For thermal neutrons ($E\sim 10^{-8}$ MeV), the cross section for 
$np\rightarrow d\gamma$ is dominated by the isovector magnetic transition 
$M1_V$ from the ${}^1S_0$ isovector channel to the ${}^3S_1$ isoscalar 
channel. At higher energies, $E\sim 1$ MeV, the cross section for 
$np\rightarrow d\gamma$ is dominated by isovector electric transitions $E1_V$ 
from the isovector $P$-wave to $^3S_1$ channel. At the energies relevant for 
BBN, the $M1_V$ and $E1_V$ transitions give comparable contributions.

Effective field theory is a useful tool in the study of physical processes 
with a clear separation of scales. This is the case for the reaction 
$np\rightarrow d\gamma$. At energies relevant for BBN, the nucleon center of 
mass momentum $p\sim 30$ MeV, the deuteron binding momentum $\gamma_t\sim 45$
MeV and the inverse of the singlet channel $^1S_0$ neutron-proton scattering
length $1/a^{exp}\sim 8$ MeV are all much smaller than the mass of the
lightest meson, the pion with 
mass $m_\pi\sim 140$ MeV. Thus a low energy EFT  can be constructed 
for momenta much below the pion mass $m_\pi$, 
 by integrating out
 the pions and other heavier degrees of freedom from the theory. The 
strong interaction of the nucleons is then described by four-nucleon local  
operators~\cite{KSW1,CRS1}. The effects of the particles that were integrated 
out of the theory are encoded in a perturbative expansion of local operators, 
where the expansion parameter is expected to be $Q/m_\pi$ with $p$, 
$\gamma_t$, $1/a^{exp}\sim Q$. The perturbative description of 
the low energy physics then allows a systematic estimation of errors at any order in the perturbation.

Recently there has been much discussion~\cite{CS,KSW1,CRS1,PRS} about the role
 EFT can play in improving the 'traditional' results obtained using 
effective-range theory (ERT) for low energy observables in the two nucleon 
system. It was shown in Ref.~\cite{bira} that for nucleon-nucleon scattering, 
ERT and low energy EFT are equivalent. In some processes, however, involving
external currents, e.g. electron-deuteron 
scattering, the ERT amplitude differs from the EFT result due to the absence of
two-body current operators. 
The most general set of allowed  multi-nucleon-external-field operators
contains operators that need not  
be related to the nucleon-nucleon scattering operators,
 included in ERT, by gauging the derivatives through minimal photon coupling.
Including 
multi-nucleon-external-field operators, e.g. two-body currents, 
is straightforward in EFT. 
 The deviation of the ERT result from EFT due to the absence of certain  
two-body current operators  
 will be greater if these operators 
appear at lower order in the perturbative EFT expansion. For example, in the
 EFT without dynamical pions a four-nucleon-one-photon operator contributes to
 deuteron quadrupole moment $\mu_Q$ at next-to-leading order 
(NLO)~\cite{CRS1}. The absence of such an operator in the ERT could be 
responsible for potential models' under prediction of $\mu_Q$ by about 
$5\%$~\cite{CRS1,PC} in the impulse approximation (IA)~\cite{SKTS,WSS}.  
In $n p\rightarrow d\gamma$, the $M1_V$
 transition amplitude also involves a four-nucleon-one-photon operator at NLO 
in the EFT that is not included in the ERT. This operator contributes about 
$5\%$
 to the cold neutron capture
 cross section $\sigma^{exp}=334.2$ mb and about $2\%$ at energy 
$E\sim 0.5$ MeV~\footnote{We would like to mention that for thermal neutrons, 
meson exchange 
currents explain the $10\%$ discrepancy in the cross section for 
$np\rightarrow d\gamma$ between potential models 
in IA and
$\sigma^{exp}$~\cite{RB}. However, modern potential models,
 e.g. Argonne $v_{18}$, with  
meson exchange currents and relativistic corrections underpredict 
 $\mu_Q$ by about $4\%$~\cite{WSS}.}.
 On the other hand, the $E1_V$ amplitude can be written 
entirely in terms of nucleon-nucleon scattering operators up to N$^3$LO and 
reduces to the ERT result at this order.

The $M1_V$ transition amplitude has been calculated up to 
NLO~\cite{CS,SSW,CRS2}. The unknown coupling $L_1^{M1}$, associated with a 
four-nucleon-one-photon operator, appearing at this order can be determined 
from the cold neutron capture rate of $\sigma^{exp}=334.2\pm 0.5$ 
mb~\cite{CWC} at incident neutron speed $v=2200$ m/s. The $E1_V$ transition
 amplitude has been calculated up to N$^3$LO~\cite{CS}. In this  work, we 
calculate the $M1_V$ transition amplitude up to N$^2$LO and the $E1_V$ 
transition amplitude up to N$^4$LO. For the $M1_V$ amplitude, there is a new 
unknown coupling $L^{M1_V}_3$ at N$^2$LO. Only a linear combination of 
$L^{M1_V}_1$ and $L^{M1_V}_3$ contributes at very low momentum. We derive 
perturbative constrains~\cite{CRS2,RS} on these couplings to reproduce the 
low energy cross section $\sigma^{exp}$. For the $E1_V$ transition, there is a
 coupling $L^{E1_V}_1$ at N$^4$LO that is not determined from nucleon-nucleon
 scattering data. We determine $L^{E1}_1$ from a $\chi^2$ 
fit to data~\cite{book:AS} for the related process of deuteron
photodisintegration  $\gamma d\rightarrow np$.

The organization of the paper is as follows: We first describe the relevant
Lagrangian and the power counting rules in Section~\ref{section:lagranian}. 
This section is rather technical and primarily used to define various 
parameters that enter the expression for the cross section. The calculation of
 the total cross section is presented in Subsections~\ref{subsection:E1V},
~\ref{subsection:M1V} and~\ref{subsection:E2M1S}. The renormalization group
 (RG) flow of the couplings
 is discussed in some detail, and from the analysis, constraints on some of 
the couplings are derived. In Subsection~\ref{subsection:fit}, all the
 remaining couplings are determined. In Section~\ref{section:results}, we
tabulate the calculated cross section 
 for some energies relevant for BBN, discuss the theoretical errors, and
 compare our results with the corresponding values from the on-line ENDF/B-VI 
database~\cite{ENDF}. Summary and 
 conclusions follow in Section~\ref{section:conclusions}.

\end{section}

\begin{section}{The Lagrangian and Power Counting}
\label{section:lagranian}
A perturbative EFT calculation requires a Lagrangian, and a set of power
counting rules  
that determine the relative sizes of diagrams contributing to a physical 
process. The nucleon kinetic energy term is
\beq\eqn{Lagrangian:kinetic}
{\mcal L}_1\Align N^\dagger\[iD_0+\frac{{\bf D}^2}{2 M_N}
-\frac{D_0^2}{2 M_N}\]N\ ,
\eeq
where the covariant derivative is
\beq
D^\mu\Align \partial^\mu +ie\frac{1+\tau_3}{2} A^\mu\ \cdot
\eeq
$N$ is an isodoublet field representing the nucleons, 
the matrix $\tau_3$ acts on the nucleon isospin space 
and $M_N=938.92$ MeV is the isospin averaged value of the nucleon mass. 
 Note that the $D_0^2$ term in 
Eq.~\refeq{Lagrangian:kinetic} includes all the relativistic corrections to the
nucleon kinetic energy~\cite{CRS1}.

The two-body Lagrangian relevant for radiative capture process $np\rightarrow
d\gamma$  
and the photodisintegration of the deuteron $\gamma d\rightarrow np$ can be 
divided into (a) a Lagrangian ${\mcal L}_2^{S}$ contributing to 
nucleon-nucleon 
scattering in the $^3S_1$ and $^1S_0$ channels, (b) a Lagrangian 
${\mcal L}_2^{P}$ contributing to  nucleon-nucleon scattering in relative 
angular momentum states $^3P_0$, $^3P_1$, $^3P_2$ and (c) a Lagrangian 
${\mcal L}_2^{EM}$ describing two-body currents that are not contained in the 
previous two Lagrangians.

\begin{subsection}{Lagrangian for $S$-State Nucleon-Nucleon Interaction:
 ${\mcal L}_2^{S}$}
\label{subsection:L3S1}
We start by describing the Lagrangian contributing to the final or initial 
state nucleon-nucleon interaction responsible for binding the deuteron. In the
 $^3S_1$
 channel, up to N$^4$LO:
\beq\eqn{Lagrangian:3S1} 
{\mcal L}_2^{(^3S_1)}\Align - C_0 (N^T P_i N)^\dagger (N^T P_i N)+
 \frac{C_2}{2}\[(N^T{\mcal O}_i^{(2)}N)^\dagger(N^T P_i N)+h.c.\]\nonumber\\
&{}& -C_4 (N^T {\mcal O}_i^{(2)} N)^\dagger (N^T {\mcal O}_i^{(2)}N)
-\frac{\CC_4}{2}\[(N^T{\mcal O}_i^{(4)}N)^\dagger(N^T P_i N)+ h.c.\]\nonumber\\
&{}&+\frac{C_6}{2}\[(N^T{\mcal O}_i^{(4)} N)^\dagger(N^T {\mcal O}_i^{(2)}N)
+h.c.\]+\frac{\CC_6}{2}\[(N^T{\mcal O}_i^{(6)}N)^\dagger(N^T P_i N)+h.c.\]\\
&{}&-C_8^\alpha(N^T {\mcal O}_i^{(4)}N)^\dagger(N^T {\mcal O}_i^{(4)}N)
-\frac{C_8^\beta}{2}\[(N^T {\mcal O}_i^{(6)}N)^\dagger(N^T {\mcal O}_i^{(2)}N)
+h.c.\]\nonumber\\
&{}&+C_{2}^{(sd)}\[(N^T P_i N)^\dagger (N^T{\mcal O}^{x y j}N){\mcal T}^{i j x
 y}+h.c.\]\nonumber\ ,
\eeq 
where summation over the repeated indices is implied. 
The $P_i$ matrices are used to project onto the ${}^3S_1$ state,
\beq
P_i\equiv\frac{1}{\sqrt{8}}\sigma_2\sigma_i\otimes\tau_2\ ,\hspace{.5in} 
{\rm Tr}[{P_i}^\dagger P_j]=\frac{1}{2}\delta_{i j}\ ,
\eeq
where the $\sigma$ matrices act on the nucleon spin space and the $\tau$ 
matrices act on the nucleon isospin space. The Galilean invariant covariant 
derivative operators are defined as:
\beq\eqn{derivatives} 
{\mcal O}^{(2)}_i\Align\frac{1}{4}\[\DL^2 P_i -2\DL P_i\DR+P_i\DR^2\]\nonumber
\\
{\mcal O}^{(4)}_i\Align\frac{1}{16}\[\DL^4P_i-4\DL^3 P_i\DR+6\DL^2 P_i\DR^2
-4\DL P_i\DR^3+P_i\DR^4\]\\
{\mcal O}^{(6)}_i\Align\frac{1}{64}\[\DL^6 P_i -6\DL^5 P_i\DR 
+ 15\DL^4 P_i\DR^2 - 20\DL^3 P_i\DR^3 \right.\nonumber\\
&{}&\left.+ 15\DL^2 P_i\DR^4 - 6\DL P_i\DR^5 + P_i\DR^6\]\nonumber\\
{\mcal O}^{x y j}\Align\frac{1}{4}\[\DL^x\DL^y P_j -\DL^x P_j \DR^y 
- \DL^y P_j \DR^x +P_j\DR^x\DR^y\]\nonumber\\
{\mcal T}^{i j x y}\Align\(\delta^{i x}\delta^{j y}
-\frac{1}{n-1}\delta^{i j}\delta^{x y}\)\nonumber\ ,
\eeq
where $n$ is the number of space-time dimensions. For the $M1_V$ transition 
there is also initial or final state nucleon-nucleon interaction in the 
${}^1S_0$ channel and the Lagrangian has the same form as the one described in
 Eq.~\refeq{Lagrangian:3S1}, with the corresponding projections onto the 
$^1S_0$ channel. Note that there is no corresponding $S$-$D$ mixing term, 
$C_2^{(sd)}$, in the $^1S_0$ channel.

 In 1997, Kaplan, Savage and Wise (KSW) formulated a  
systematic power counting for calculating the cross sections of two-nucleon
processes  
in the context of EFT~\cite{KSW1}. One key ingredient in the KSW scheme is the 
non-perturbative treatment of the large $S$-channel nucleon-nucleon scattering
length. KSW power counting can be implemented using power divergence 
subtraction (PDS) where one subtracts the loop-divergences in both $n=3$ and
$n=4$ space-time dimensions. An alternative to PDS is off-shell
subtraction,   
where the divergences are regulated by subtracting the amplitude at an 
off-shell momentum point~\cite{GMS}. In this paper, we will use PDS scheme.

The KSW power counting is as follows: The expansion parameter is
$Q/\Lambda$. The 
 nucleon center of mass momentum $p$, the deuteron binding momentum 
$\gamma_t$, the inverse of the $^1S_0$ channel scattering length $1/a^{exp}$
  and the
 renormalization scale $\mu$  are formally considered ${\mcal O}(Q)$ 
and $\Lambda\sim m_\pi/2$ for 
this low energy EFT (the factor of half comes from the analytic structure of 
the one-pion-exchange contributions). The couplings $C_{2n}$ scale as 
$1/Q^{n+1}$, $\CC_{2n}$ scale as $1/Q^{n}$ and $C_2^{(sd)}$ scales as $1/Q$. 
For the low energy theory, we formally take $m_\pi/M_N\sim Q/\Lambda$. Thus,
relativistic corrections which come in as $p^2/M_N^2$, 
$\gamma_t^2/M_N^2\sim Q^2/M_N^2=Q^2/m_\pi^2\times m_\pi^2/M_N^2$ 
contribute at N$^4$LO.

It is a feature of the KSW power counting that the EFT couplings associated 
with $S$-state interactions scale with some powers of $Q$. This is because the 
couplings are fine-tuned to reproduce the large scattering lengths that one 
sees in the $^1S_0$ and $^3S_1$ channels. In order to reproduce the exact 
deuteron pole, the coefficients $C_{2n}$ are expanded in powers of $Q$:
\beq
C_0\Align C_{0,-1}+C_{0,0}+\cdots\nonumber\\
C_2\Align C_{2,-2}+C_{2,-1}+\cdots\\
\vdots&&\nonumber
\eeq
where the second subscript denotes the $Q$ scaling. The EFT couplings 
$C_{2n}$ are determined by matching the nucleon-nucleon
 scattering 
amplitude calculated in EFT~\cite{KSW1,CRS1,RS} and that obtained from the
ERE\cite{schwinger}: 
 
\beq\eqn{ERE}
p \cot\delta\Align -\gamma_t+\frac{1}{2}\rho_d\(p^2+\gamma_t^2\)
+\frac{1}{2} w_2 \(p^2+\gamma_t^2\)^2 +\cdots
\eeq
To the order we are 
working, the deuteron binding momentum $\gamma_t\approx\gamma-\gamma^3/(8
M_N^2)$,  
where $\gamma=\sqrt{M_N B}$, and 
$B=2.2246$ MeV~\cite{STS} is the deuteron binding energy. 
The coupling $C_{2}^{(sd)}=3\eta_d/(\sqrt{2}\gamma^2)$ is determined from the 
$S$-$D$ mixing parameter $\overline\epsilon_1$~\cite{CRS1,CRS2,PKR}, where 
$\eta_d=0.02534$ is the $D$-wave to $S$-wave ratio at the deuteron 
pole~\cite{STS}. Notice that $C_{2}^{(sd)}$ is smaller than naive power 
counting estimate because of the smallness of the parameter $\eta_d$.

For our calculation, we need the $^3S_1$ channel initial or final state 
interaction up to N$^4$LO and the $^1S_0$ interaction up to N$^2$LO.  In the 
$^3S_1$ channel, at N$^4$LO only four experimental inputs ($\gamma$, 
$\rho_d$, $w_2$ and $\eta_d$) enter the scattering amplitude. The EFT 
couplings that depend on the high energy scale can all be expressed in terms 
of these four parameters. However, only the combinations $C_4+\CC_4$ and 
$C_{8}^\alpha +C_{8}^\beta\equiv C_8$ contribute to nucleon-nucleon scattering
and so nucleon-nucleon scattering data cannot be used to separate $C_4$ from
$\CC_4$, and $C_{8}^\alpha$ from $C_{8}^\beta$.
We use the effective range $\rho_d=1.764$ fm and shape parameter 
$w_2=0.778$ fm$^3$\cite{STS}. In the $^1S_0$ channel, at N$^2$LO the 
nucleon-nucleon scattering amplitude is completely determined by the 
scattering length $a^{exp}$ and the effective range $r_0$. The EFT couplings
 $C_{2n}$ are determined in terms of these two parameters. 
Note that the experimentally
 measured scattering length $a^{exp}$, for neutron-proton scattering in the
 singlet channel ${}^1S_0$, gets about a $3\%$ contribution from 
magnetic moment interaction~\cite{WSS}. Isospin-breaking single magnetic 
photon exchange
 gives the dominant correction and it is described by the Lagrangian
\beq\eqn{isospin}
{\mcal L}=\frac{e\kappa_1}{2 M_N}N^\dagger\tau_3\ {\bf \sigma\cdot B} N\ ,
\eeq
where ${\bf B}$ is the magnetic field and
$\kappa_1=2.353$ is the isovector nucleon magnetic moment in nuclear 
magnetons. These N$^2$LO interactions must be included in the $M1_V$ 
amplitude. We implicitly include the magnetic moment interactions by 
using the experimental value $a^{exp}=-23.75$ fm~\cite{KN}.
For the effective range we take $r_0=2.73$ fm. 
\end{subsection}

\begin{subsection}{Lagrangian for $P$-State Nucleon-Nucleon Interaction:
 ${\mcal L}_2^{P}$}
\label{subsection:L2P}
In this subsection,
the second Lagrangian ${\mcal L}_2^{P}$ is described. For the $E1_V$ 
transition amplitude, there are also initial or final state nucleon-nucleon 
interactions in relative angular momentum states 
${}^3P_0,\ {}^3P_1, {}^3P_2$ described by the Lagrangian~\cite{CS}:
\beq\eqn{Lagrangian:3PN}
{\mcal L}_2^P\Align \( C_{2}^{(3P_0)}\delta_{x y}\delta_{w z}
+C_{2}^{(3P_1)}\[\delta_{x w}\delta_{y z}-\delta_{x z}\delta_{y w}\]
+C_{2}^{(3P_2)}\[2\delta_{x w}\delta_{y z}+2\delta_{x z}\delta_{y w}
-\frac{4}{3}\delta_{x y}\delta_{w z}\] \)\nonumber\\
&{}&\hspace{1in}\times\frac{1}{4} (N^T {\mcal O}_{x y}^{(1,P)} N)^\dagger(N^T
 {\mcal O}_{w z}^{(1,P)} N)\ ,
\eeq
where the $P$-wave operator is
\beq
 {\mcal O}_{i j}^{(1,P)}
\Align\stackrel{\leftarrow}{\bf D}_i P_j^{(P)}-P_j^{(P)}
\stackrel{\rightarrow}{\bf D}_i\ ,
\eeq
and $P_i^{(P)}$ are the spin-isospin projectors for the isotriplet, 
spintriplet channel
\beq
P_i^{(P)}\equiv\frac{1}{\sqrt{8}}\sigma_2\sigma_i\ \tau_2\tau_3\ ,
\hspace{.2in}{\rm Tr}\[P_i^{(P)}P_j^{(P)}\]=\frac{1}{2}\delta_{i j}\ \cdot
\eeq
The power counting of the $P$-state couplings is straightforward. Since there 
are no fine tuned high energy scales in the $P$-state scattering, the  
couplings that are dependent only on the high energy physics are 
${\mcal O}(1)$ i.e. they do not scale with $Q$. As a result, the next set of
$P$-state operators,  
suppressed by two extra powers of momentum $p$, enters only at N$^5$LO. Only 
the linear combination
\beq
C_p&\equiv&C_2^{(^3P_0)}+2C_2^{(^3P_1)}+\frac{20}{3} C_2^{(^3P_2)}
\eeq
enters our calculation and Nijmegen phase shift analysis~\cite{SKTS,SKRS}
fixes $C_p=-1.49$ fm$^4$~\cite{CS}.

\end{subsection}

\begin{subsection}{Lagrangian for Two-Body Currents: ${\mcal L}_2^{EM}$}
\label{subsection:LEM}
Finally, we discuss the two-body currents contributing to 
$np\rightarrow d\gamma$ or $\gamma d\rightarrow np$, which are not included in
the previous two Lagrangians ${\mcal L}_2^S$ and ${\mcal L}_2^P$.
 These operators are not 
related by gauge transformation to the nucleon-nucleon scattering operators in 
${\mcal L}_2^S$ and ${\mcal L}_2^P$.
\beq\eqn{Lagrangian:currents}
{\mcal L}_2^{EM}
\Align e\,L^{M1_V}_1(N^T P_i N)^\dagger(N^T \overline{P}_3 N){\bf B}_i 
-e\,L^{M1_V}_3(N^T {\mcal O}_i^{(2)} N)^\dagger(N^T \overline{P}_3 N){\bf B}_i
\nonumber\\
&&-e\,{\tilde L}^{M1_V}_3(N^T P_i N)^\dagger(N^T {\overline{\mcal O}}_3^{(2)}
 N){\bf B}_i 
+\frac{1}{2}e\,L^{E1_V}_1(N^T{\mcal O}_{i a}^{(1,P)} N)^\dagger(N^T P_a N)
{\bf E}_i\nonumber\\
&{}&-\frac{1}{2}e\,L^{E1_V}_3(N^T {\mcal O}_{i a}^{(1,P)}N)^\dagger(N^T{\mcal 
O}^{(2)}_aN){\bf E}_i 
-\frac{1}{2}e\,{\tilde L}^{E1_V}_3(N^T{\mcal O}_{i a}^{(3,P)} N)^\dagger(N^T 
P_aN){\bf E}_i\\
&{}&+\frac{1}{2}e\, L^{E1_V}_5(N^T {\mcal O}_{i a}^{(1,P)}N)^\dagger(N^T 
{\mcal O}_a^{(4)}N){\bf E}_i +\frac{1}{2}e\,{\tilde L}_5^{E1_V}(N^T {\mcal O}_
{i a}^{(3,P)}N)^\dagger(N^T {\mcal O}_a^{(2)}N){\bf E}_i\nonumber\\
&{}&-e\,L_2^{M1_S}i\epsilon_{ijk}(N^T P_iN)^\dagger (N^T P_j N) {\bf B}_k
+ h.c. \nonumber\ ,
\eeq
where ${\bf E}$ and ${\bf B}$ are the electric and magnetic fields 
respectively, and  
\beq
\overline{P}_i\Align\frac{1}{\sqrt{8}}\sigma_2\otimes\tau_2\tau_i\nonumber\\
\overline{\mcal O}^{(2)}_i\Align\frac{1}{4}\[\DL^2 \overline{P}_i -2\DL 
\overline{P}_i\DR+\overline{P}_i\DR^2\]\\
{\mcal O}_{i j}^{(3,P)}\Align\frac{1}{4}\[\DL_i\DL^2 P_j^{(P)}-2\DL_i\DL_k 
P_j^{(P)}\DR_k+\DL_i P_j^{(P)}\DR^2\right.\nonumber\\
&{}&\left.-\DL^2 P_j^{(P)}\DR_i+2\DL_k P_j^{(P)}\DR_k\DL_i
- P_j^{(P)}\DR^2\DR_i\]\nonumber\ \cdot
\eeq
The superscript on the two-body current couplings $L$'s denote the transitions 
that the particular operators contribute to. For the $M1_V$ transition only a 
specific $p$ independent combination of $L_1^{M1_V}$ and $L_3^{M1_V}$ 
contributes and we fix it from the cold neutron capture rate 
$np\rightarrow d\gamma$. Renormalization group (RG) flow analysis determines  
${\tilde L}\!^{M1_V}_3$ in terms $L^{M1_V}_1$, $C_2^{(^1S_0)}$ and 
$C_2^{(^3S_0)}$. For the $E1_V$ transition, only a combination, $L_{E1}$, of 
$L_1^{E1_V}$, 
$L_3^{E1_V}$ and $L_5^{E1_V}$ enters the calculation at N$^4$LO and it is 
fixed from a $\chi^2$ fit to the related deuteron breakup process 
$\gamma d\rightarrow np$~\cite{CS}. The other operators contribute at orders 
higher than considered here. The RG analysis of these operators is discussed 
in more detail below, in
Subsections~\ref{subsection:E1V},~\ref{subsection:M1V}. Finally, the observed
value of the deuteron magnetic  
moment $\mu_M$ fixes the isoscalar magnetic moment coupling 
$L_{2}^{(M1_S)}=-0.149$ fm$^4$~\cite{CRS1,KSW2}.  

\end{subsection}
In the last three subsections, we described the Lagrangian and the power
 counting rules relevant to our calculation. Now, the calculation for the
 total cross section is presented, along with the estimated theoretical 
uncertainty. This is followed by the discussion of a matching procedure for 
determining the unknown couplings $L_1^{M1_V}$ and $L_3^{M1_V}$. The 
parameter $L_{E1}$ is fixed from the $\gamma d\rightarrow np$ data. 

\end{section}

\begin{section}{Calculation and Analysis}
\label{calculation}
The amplitude $\Gamma$ for low energy $np\rightarrow d\gamma$ reaction
is~\cite{SSW,CRS2}: 
\beq\eqn{Amplitude}
i\ \Gamma\Align e\ X_{E1_V} U^T \tau_2\tau_3\ \sigma_2\
\sigma\cdot\epsilon_{(d)}^\ast U\ p\cdot\epsilon_{(\gamma)}^\ast +ie\ X_{M1_V}
\varepsilon^{abc}\epsilon_{(d)}^{\ast a} \bfk^b\epsilon_{(\gamma)}^{\ast c}\
U^T\tau_2\tau_3\ \sigma_2 U\nonumber\\ 
&{}&+e\ X_{M1_S}\frac{1}{\sqrt{2}}U^T\tau_2\ \sigma_2\[\sigma\cdot\bfk\
\epsilon_{(d)}^\ast\cdot\epsilon_{(\gamma)}^\ast-\epsilon_{(d)}^\ast\cdot\bfk\
\epsilon_{(\gamma)}^\ast\cdot\sigma\] U\\ 
&{}&+e\ X_{E2_S}\frac{1}{\sqrt{2}}U^T\tau_2\ \sigma_2\[\sigma\cdot\bfk\
\epsilon_{(d)}^\ast\cdot\epsilon_{(\gamma)}^\ast+\epsilon_{(d)}^\ast\cdot\bfk\
\epsilon_{(\gamma)}^\ast\cdot\sigma-\frac{2}{n-1}\sigma\cdot\epsilon_{(d)}^\ast
\bfk\cdot\epsilon_{(\gamma)}^\ast\] U\nonumber\ , 
\eeq
where only the lowest partial waves are shown: isovector electric dipole 
capture of nucleons in a $P$-wave with amplitude $X_{E1_V}$, isovector 
magnetic dipole capture of nucleons in ${}^1S_0$ state with amplitude
$X_{M1_V}$,  
isoscalar magnetic dipole capture of nucleons in  ${}^3S_1$ state with 
amplitude
$X_{M1_S}$ and isoscalar electric quadrupole capture of nucleons with 
amplitude $X_{E2_S}$. We use dimensional regularization to regulate 
divergences and $n$ represents the number of space-time dimensions. The $U$'s 
are two component nucleon spinor wave functions. 
$|\bfp|\equiv p$ is the nucleon center 
of mass momentum, $k$ is the outgoing photon momentum, and 
$\epsilon_{(\gamma)}$ and $\epsilon_{(d)}$ are the photon and deuteron 
polarization vectors respectively. The following dimensionless amplitudes, 
${\tilde X}$, are defined
\beq
\frac{|\bfp| M_N}{\gamma^2+p^2}X_{E1_V}\Align
i\frac{2}{M_N}\sqrt{\frac{\pi}{\gamma^3}}{\tilde X}_{E1_V},\hspace{.2in}
X_{M1_V}=i\frac{2}{M_N}\sqrt{\frac{\pi}{\gamma^3}}{\tilde X}_{M1_V}\nonumber\\ 
 X_{M1_S}\Align i\frac{2}{M_N}\sqrt{\frac{\pi}{\gamma^3}}{\tilde
X}_{M1_S},\hspace{.2in}
X_{E2_S}=i\frac{2}{M_N}\sqrt{\frac{\pi}{\gamma^3}}{\tilde X}_{E2_S}\ , 
\eeq
 and the total cross section for $np\rightarrow d\gamma$ is then written as:
\beq\eqn{crosssection}
\sigma\Align 4\alpha\pi\(1-\frac{2{p^4}+4{p^2}{{\gamma }^2}+3{{\gamma }^4}}
    {4{M_N^2}({p^2}+{{\gamma }^2})}\)\frac{\(\gamma^2+p^2\)^3}{\gamma^3 M_N^4
 p}\ \frac{1}{2}\int\limits_{-1}^{1}d\cos\theta\[\frac{3}{2}|{\tilde X}_{E1_V}
|^2 \sin^2\theta+|{\tilde X}_{M1_V}|^2\right.\nonumber\\
&&\hspace{4in}\left.+|{\tilde X}_{M1_S}|^2+|{\tilde X}_{E2_S}|^2\frac{}{}\]\\
&\equiv& 4\alpha\pi\(1-\frac{2{p^4}+4{p^2}{{\gamma }^2}+3{{\gamma }^4}}
    {4{M_N^2}({p^2}+{{\gamma }^2})}\)
\frac{\(\gamma^2+p^2\)^3}{\gamma^3 M_N^4 p}\[\langle{\tilde X}_{E1_V}\rangle^2
+\langle{\tilde X}_{M1_V}\rangle^2+\langle{\tilde X}_{M1_S}\rangle^2
+\langle{\tilde X}_{E2_S}\rangle^2\]\nonumber\ ,
\eeq
where the deuteron mass is $M_d\approx 2 M_N-\gamma^2/M_N$~\cite{CRS1}. The 
relativistic corrections enter at N$^4$LO and so only the $E1_V$ cross section
 which is calculated to this order receives such contributions.

\begin{subsection}{Isovector Electric Transition: $E1_V$}
\label{subsection:E1V}
The  $E1_V$ amplitude up to N$^4$LO is
\beq\eqn{E1Vrho}
{\tilde X}_{E1_V}\!\!\Align -\frac{M_N
p\gamma^2}{(p^2+\gamma^2)^2}\[1+\frac{\rho_d\gamma}{2}
+\frac{3}{8}\rho_d^2\gamma^2+\frac{5}{16}\rho_d^3\gamma^3
+\frac{35}{128}\rho_d^4\gamma^4-\frac{5}{2}\eta_d^2+\frac{(p^2+\gamma^2)^2}{2
M_N} L_{E1}\right.\nonumber\\ 
&&\hspace{0.5in}+\frac{M_N\gamma}{12\pi}\(\frac{\gamma^2}{3}+p^2\)C_p
\(1+\frac{\rho_d\gamma}{2}\)+\frac{|p|}{M_N}\cos\theta
\(1+\frac{\rho_d\gamma}{2}+\frac{3}{8}\rho_d^2\gamma^2\)\\
&&\left.\hspace{0.5in}+\frac{p^2}{M_N^2}\cos^2\theta
-\frac{8{p^4}+17{p^2}{{\gamma }^2}+5{{\gamma }^4}}
    {16{M_N^2}({p^2}+{{\gamma }^2})}\]\nonumber\ ,
\eeq
where the relativistic corrections contribute at 
N$^4$LO as they are suppressed by additional powers of $(m_\pi/M_N)^2$, and 
\beq\eqn{LE1}
L_{E1}&\equiv&\frac{L_{1,-1}^{E1}-\gamma^2L_{3,-3}^{E1}+\gamma^4
L_{5,-5}^{E1}+\gamma^4
M_N(C_{8,-5}^\beta+2C_{8,-5}^\t)-M_N{_\eta\hspace{-.05cm}}\CC_{4,-1}}
{C_{0,-1}^\t}\\
{_\eta\hspace{-.05cm}}\CC_{4,-1}&\equiv&\CC_{4,-1}^\t
+\frac{2\pi}{M_N}\frac{\rho_d w_2\gamma^2}{(\mu-\gamma)^3}\nonumber\ \cdot
\eeq
The ${_\eta\hspace{-.05cm}}\CC_{4,-1}$ coupling renormalizes contributions
proportional to $\eta_d^2$ in Eq.~\refeq{E1Vrho}.
The renormalization scale $\mu$ independent parameter $L_{E1}\sim {\mcal
O}(1)$ can be simplified 
further. We assume that all the couplings have a sensible large scattering 
length ($\gamma\rightarrow 0$) limit. Then it follows that at the high scale
$\mu=\Lambda$, only the combination
\small{$ (L_{1,-1}^{E1}-M_N{_\eta\hspace{-.05cm}}\CC_{4,-1})/C_{0,-1}^\t$}
 \normalsize contributes a 
${\mcal O}(1)$ 
term. The other combinations, e.g. 
\small{$\gamma^2L_{3,-3}^{E1}/C_{0,-1}^\t$}\normalsize ,  
contribute terms of ${\mcal O}(Q^2)$ and higher. Thus RG analysis gives
\beq
L_{E1}\equiv\frac{L_{1,-1}^{E1}-M_N{_\eta\hspace{-.05cm}}\CC_{4,-1}}
{C_{0,-1}^\t}\(1+{\mcal O}(Q^2)\), 
\eeq
and we keep only the leading contribution to $L_{E1}$ for our calculation. 
It is possible to make an order of magnitude estimate for $L_{E1}$. 
The contribution from $S$-$D$ mixing,  proportional 
to $\eta_d^2$, is numerically negligible 
at N$^4$LO even though it is formally a N$^4$LO term.
Thus, ignoring contribution from ${_\eta\hspace{-.05cm}}\CC_{4,-1}$, we 
estimate at the renormalization scale $\mu=\Lambda\sim m_\pi/2$:
\beq\eqn{estimateLE1}
|C_{0,-1}^\t|&\sim&\frac{4\pi}{M_N}\frac{1}{\Lambda}, \hspace{0.5in}
|L_{1,-1}^{E1}|\sim \frac{1}{M_N \Lambda^4}
\nonumber\\
\Rightarrow &&|L_{E1}|\sim 2\ {\rm fm}^3,
\eeq
where factors of $1/M_N$ from the nucleon loops were included. 

In Eq.~\refeq{E1Vrho}, we have only kept the terms that contribute to the 
total cross section after summing over deuteron, photon and nucleon 
polarizations. We have included 
the formally N$^2$LO contribution proportional to $\cos\theta$. This term, odd
 in $\cos\theta$, contributes to the cross section only at N$^4$LO after the 
integration over the angle $\theta$, Eq.~\refeq{crosssection}. In obtaining 
Eq.~\refeq{E1Vrho}, we have also used the fact that only the $\mu$ 
independent combination
\beq
\frac{{\tilde L}^{(E1_V)}_{3,-3}-\gamma^2{\tilde L}^{(E1_V)}_{5,-5}-2 M_N
\CC_{6,-3}^\t+2 M_N \gamma^2 C_{8,-5}^\beta}{C_{0,-1}^\t} 
\eeq
enters our result. From RG analysis, by flowing the couplings to the high 
scale $\mu=\Lambda$, we see that the numerator is ${\mcal O}(1/Q)$ instead of 
the naive counting of $1/Q^3$. Thus this particular combination of operators 
contributes starting at N$^6$LO and we ignore it in this N$^4$LO calculation. 

 The dominant contribution beyond leading order, Eq.~\refeq{E1Vrho}, is a 
simple expansion of the factor $\sqrt{Z_d}\equiv 1/\sqrt{1-\rho_d\gamma}$ from
 the deuteron wave function renormalization
\beq\eqn{wavefunction}
{\mcal
Z}\Align-\frac{8\pi\gamma}{M_N^2}\[1+\rho_d\gamma+\rho_d^2\gamma^2
+\rho_d^3\gamma^3
+\rho_d^4\gamma^4+\eta_d^2\frac{7\gamma-5\mu}{\mu-\gamma}+\frac{\gamma^2}{8
M_N^2}\ \frac{7\mu-5\gamma}{\mu-\gamma}+\cdots\]\ \cdot 
\eeq
In Eq.~\refeq{wavefunction}, the $\rho_d\gamma$ terms inside the square
brackets are from the expansion of $Z_d$, and the sixth and seventh terms arise
from the $S$-$D$ mixing and relativistic corrections to the deuteron two-point
function respectively.  
 
 Amplitudes for inelastic processes, e.g. $E1_V$ and $M1_V$ transitions, with 
a deuteron in the final state involve a factor of $\sqrt{Z_d}$. On the other
 hand amplitudes for elastic processes involving a deuteron, e.g. the deuteron
 quadrupole form factor $F_Q$,  include a factor of $Z_d$. In 
Ref.~\cite{PRS}, an expansion in $Z_d -1$ was proposed which resums the 
expansion in $\rho_d\gamma$ seen in Eq.~\refeq{wavefunction} and thereby 
reproduces the exact factor of $Z_d$ at NLO for the amplitude for elastic 
processes. On the other hand, amplitudes involving inelastic processes on the
 deuteron have this sum incomplete. The factor of $\sqrt{Z_d}$ is reproduced 
only in perturbation. This incomplete sum can be avoided if we do an expansion
 in $\sqrt{Z_d}-1$ to reproduce the exact factor of $\sqrt{Z_d}$ for inelastic
 processes. The overall factor of $Z_d$ in elastic processes is then 
reproduced at N$^2$LO instead of NLO. Thus, one might think that the 
$\sqrt{Z_d}-1$ expansion is more appropriate to calculate the cross section 
for the inelastic process $np\rightarrow d\gamma$. However, for the total 
cross section, which is summed over the nucleon spins and the deuteron and 
photon polarizations, it is the square of the amplitudes that enter the 
expression. In such cases it is more convenient to do the expansion in $Z_d-1$
 and reproduce the exact factor of $Z_d$ at NLO. 
The three different expansions $\rho_d\gamma$, $Z_d-1$, $\sqrt{Z_d}-1$ are
 formally equivalent but correspond to different ways of relating the $C_{2n}$
 couplings to the ERE. 

In the $Z_d-1$ expansion, the position of the 
deuteron pole at momentum $p=i\gamma_t\approx i\gamma$ and the residue $Z_d$ 
(without $S$-$D$ mixing) at the pole are used to fix the $C_{2n}$ couplings.
One can think of the $Z_d-1$ expansion as reproducing the
asymptotic tail of the deuteron wave function exactly. This allows a better
description of data as low energy experiments probe only the large distance
behavior of the deuteron wave function. 
For the rest of the paper we will use the this expansion. The results for
${\tilde X}$ can then be written as:   
\beq\eqn{amplitude:E1VZ}
{\tilde X}_{E1_V}\!\!\Align -\frac{M_N
p\gamma^2}{(p^2+\gamma^2)^2}\[1+\frac{1}{2}(Z_d-1)-\frac{1}{8}(Z_d-1)^2
+\frac{1}{16}(Z_d-1)^3-\frac{5}{128}(Z_d-1)^4\right.\nonumber\\
&&+\frac{M_N\gamma}{12\pi}\(\frac{\gamma^2}{3}+p^2\)C_p
\(1+\frac{1}{2}(Z_d-1)\)-\frac{5}{2}\eta_d^2+\frac{(p^2+\gamma^2)^2}{2 M_N}
L_{E1}\nonumber\\
&&\nonumber\\
&&\left.-\frac{8{p^4}+17{p^2}{{\gamma }^2}+5{{\gamma }^4}}
    {16{M_N^2}({p^2}+{{\gamma
}^2})}+\frac{|p|}{M_N}\cos\theta\(1+\frac{1}{2}(Z_d-1) 
-\frac{1}{8}(Z_d-1)^2 \)+\frac{p^2}{M_N^2}\cos^2\theta\]\nonumber\\
\langle{\tilde X}_{E1_V}\rangle^2\Align \frac{{M_N^2}{p^2}{{\gamma
}^4}}{{{({p^2}+{{\gamma }^2})}^4}} 
  Z_d\[1 +\frac{M_N\gamma }{6\pi }\(\frac{{{\gamma }^2}}{3}+{p^2}\) C_p
-\frac{5\eta_d^2}{Z_d}+\frac{{{({p^2}+{{\gamma }^2})}^2}}{Z_d M_N}L_{E1}
\right.\nonumber\\
&&\left.-\frac{8{p^4}+17{p^2}{{\gamma }^2}+5{{\gamma }^4}}
    {8 Z_d {M_N^2}({p^2}+{{\gamma }^2})}+\frac{3}{5}\frac{p^2}{Z_d M_N^2} \]\
, 
\eeq
where the last part in $\langle{\tilde X}_{E1_V}\rangle^2$ is from the
 $\cos^2\theta$ terms.

The expression for $\langle{\tilde X}_{E1_V}\rangle^2$ in
 Eq.~\refeq{amplitude:E1VZ} is a perturbative result. There are corrections to
 this expression from terms that are higher order than the ones considered 
here in the perturbative expansion. The higher order corrections can be 
separated into: (a) factors of $Z_d$ from the wavefunction renormalization of 
the deuteron, Eq.~\refeq{wavefunction} and (b) contributions from higher order
 nucleon-nucleon scattering operators and two-body currents. At 
$E\sim 1$ MeV, we expect higher order corrections to  contribute 
$(Z_d-1)(p^2+\gamma^2)/M_N^2\sim 0.004$, $5(Z_d-1)\eta_d^2\sim 0.002$ due to 
wave function renormalization. The $Z_d-1$ corrections to the term involving 
$L_{E1}$ are renormalized away when we fit $L_{E1}$. The higher
 order $P$-wave operators should contribute
 $\gamma p^4/(6 \pi\Lambda^5)$, $\gamma^3 p^2/(6 \pi\Lambda^5)$,
 $\gamma^5/(6 \pi\Lambda^5)\sim 0.01$ with  
$\Lambda\sim m_\pi/2$. Thus we estimate the theoretical error to be $\sim 1\%$
at energies of $E\sim 1$ MeV. The error is smaller at lower energies.

\end{subsection}
\begin{subsection}{Isovector Magnetic Transition: $M1_V$}
\label{subsection:M1V}
A straight forward calculation of $M1_V$ amplitude gives:
\beq\eqn{amplitude:M1V}
{\tilde X}_{M1_V}^{(0)}\Align
-\kappa_1\gamma^2\(\frac{1}{p^2+\gamma^2}-\frac{1}{(\gamma-ip)
(\frac{1}{a^{exp}}+i p)}\)\nonumber\\ 
{\tilde X}_{M1_V}^{(1)}\Align\frac{1}{2}(Z_d-1){\tilde X}_{M1_V}^{(0)}+\kappa_1
\frac{{r_0}}{2}\frac{{p^2}{{\gamma }^2}}{{{(\frac{1}{a^{exp}}+ip)}^2}(\gamma
-ip)}+\frac{\gamma^2 M_N}{4\pi(\frac{1}{a^{exp}}+ip) }\ L_{np}\\ 
{\tilde X}_{M1_V}^{(2)}\Align \(\frac{1}{2}(Z_d-1)+\MVr\)[{\tilde
X}_{M1_V}^{(1)}-\frac{1}{2}(Z_d-1){\tilde X}_{M1_V}^{(0)}]\nonumber\\ 
&& -\frac{1}{8}(Z_d-1)^2{\tilde X}_{M1_V}^{(0)}+\MVD\ {\tilde L}_{np}
\nonumber\ , 
\eeq
where the $\mu$ independent parameters are
\beq\eqn{Lnp}
L_{np}&\equiv&(\mu-\gamma)(\mu-1/a^{exp})\[L^{M1_V}_{1,-2}
-\frac{\kappa_1\pi}{M_N\gamma}\(\frac{r_0\gamma}{(\mu-\frac{1}{a^{exp}})^2}
+\frac{Z_d-1}{(\mu-\gamma)^2}\)\]\nonumber\\
{\tilde L}_{np}&\equiv&(\mu-\gamma)(\mu-1/a^{exp})\[L^{M1_V}_{1,-1}
-\gamma^2
L^{M1_V}_{3,-3}+\frac{\kappa_1\pi}{M_N\gamma}
\frac{(Z_d-1)^2}{(\mu-\gamma)^2}\]\\
\Align
(\mu-\gamma)(\mu-1/a^{exp})\[L^{M1_V}_{1,-1}
+\frac{\kappa_1\pi}{M_N\gamma}\frac{(Z_d-1)^2}{(\mu-\gamma)^2}\]\(1+{\mcal
O}(Q^2)\)\nonumber\ \cdot 
\eeq
An analysis similar to the one used for $L_{E1}$, allows us to simplify the
parameters $L_{np}$'s above in Eq.~\refeq{Lnp}. A naive estimate gives
\beq\eqn{estimateLnp}
L_{np}\sim -9\ {\rm fm}^2\ \ ,\hspace{0.5in} 
 {\tilde L}_{np}\sim  +3\ {\rm fm}^2.
\eeq
We use the experimentally measured scattering length $a^{exp}$ which incudes
 the effect of single magnetic photon exchange in neutron-proton scattering. 
In Eq.~\refeq{amplitude:M1V}, from RG analysis, we used:
\beq
{\tilde
L}^{M1_V}_{3,-3}\Align\frac{\xi}{(\mu-\gamma)(\mu-\frac{1}{a^{exp}})}
-\frac{r_0}{2}\frac{1}{\mu-\frac{1}{a^{exp}}}\[L^{M1_V}_{1,-2}
+\frac{\kappa_1\pi}{M_N\gamma}\(\frac{r_0\gamma}{(\mu-\frac{1}{a^{exp}})^2}
-\frac{Z_d-1}{(\mu-\gamma)^2}\)\]\ \cdot
\eeq
The first part ${\mcal O}(1/Q^2)$ gives higher order corrections to  
${\tilde L}\!^{M1_V}_3\sim1/Q^3$. So we can set $\xi=0$ at N$^4$LO, and have 
${\tilde L}\!^{M1_V}_3$ completely determined. Up to N$^2$LO
\beq
|{\tilde X}_{M1_V}|^2\Align|{\tilde X}_{M1_V}^{(0)}|^2+2 Re[{\tilde
X}_{M1_V}^{(0)}({\tilde X}_{M1_V}^{(1)})^\ast]+|{\tilde X}_{M1_V}^{(1)}|^2+2
Re[{\tilde X}_{M1_V}^{(0)}({\tilde X}_{M1_V}^{(2)})^\ast]\ \cdot 
\eeq

The singlet channel scattering length $a^{exp}= -23.75$ fm is unnaturally 
large and it is easier to do the error analysis for the cross section in the 
limit $a^{exp}\rightarrow\infty$, with finite corrections from 
$1/(a^{exp}\gamma)$ and $1/(a^{exp} p)$ terms. There are corrections from
 initial state interactions in the $^1S_0$ channel that come in as 
$r_0^3 p^6/\gamma^3\sim 0.03$ at center of mass energy $E=1$ MeV.
There is also a contribution from $^1S_0$
operators that come in as $r_1 p^4/\gamma\sim -0.002$, 
where the shape parameter $r_1\sim -1$ fm$^3$ 
is the $^1S_0$ equivalent of $w_2$ in Eq.~\refeq{ERE}. 
The momentum independent corrections from factors of $Z_d-1$ from deuteron 
wavefunction renormalization, $r_0\gamma$ from $^1S_0$ channel initial state 
interaction etc. get renormalized away when we fit the parameters 
$L_{np}$'s at very low momentum. 
 We estimate the errors in $M1_V$ to be $\sim 3\%$ at
 $E=1$ MeV.   

\end{subsection}
\begin{subsection}{ Isoscalar Magnetic and Electric Transitions: 
$M1_S$, $E2_S$}
\label{subsection:E2M1S}
The $M1_S$ and $E2_S$ amplitude have been calculated for cold  
neutron capture~\cite{SSW,CRS2}. The leading  contribution from these 
transitions at non-zero  
momentum transfer is:
\beq\eqn{amplitude:M1SE2S}
|{\tilde X}_{M1_S}|^2\Align\frac{{M_N^2}{{\gamma }^4}}{2{{\pi
}^2}({p^2}+{{\gamma }^2})}{(L^{M1_S}_2)}^2(\gamma -\mu)^4\nonumber\\ 
|{\tilde X}_{E2_S}|^2\Align\frac{\gamma^4
\eta_d^2}{100}\,\frac{(p^4+7p^2\gamma^2+\gamma^4)}{(p^2+\gamma^2)^4}\ \cdot 
\eeq
The coupling $L^{M1_S}_2$ can be fitted to the deuteron magnetic moment at the
 renormalization scale $\mu=m_\pi$ and  gives 
$L^{M1_S}_2(m_\pi)=-0.149$ fm$^4$~\cite{CRS1,KSW2}.

\end{subsection}

In the last three subsections, 
the $np\rightarrow d\gamma$ amplitude in terms of the lowest partial waves, 
$M1_V$, $E1_V$, $M1_S$ and $E2_S$ has been calculated in terms of three 
unknown parameters $L_{np}$, ${\tilde L}_{np}$ and $L_{E1}$,  
Eqs.~\refeq{amplitude:E1VZ},~\refeq{amplitude:M1V} 
and~\refeq{amplitude:M1SE2S}. 
We now move on to describe how the three 
 unknown parameters $L_{np}$, ${\tilde L}_{np}$ and $L_{E1}$ can be determined 
from $np\rightarrow d\gamma$ for cold neutrons
 and from photodisintegration of the deuteron $\gamma d\rightarrow np$. 

\makefig{compare.fig}{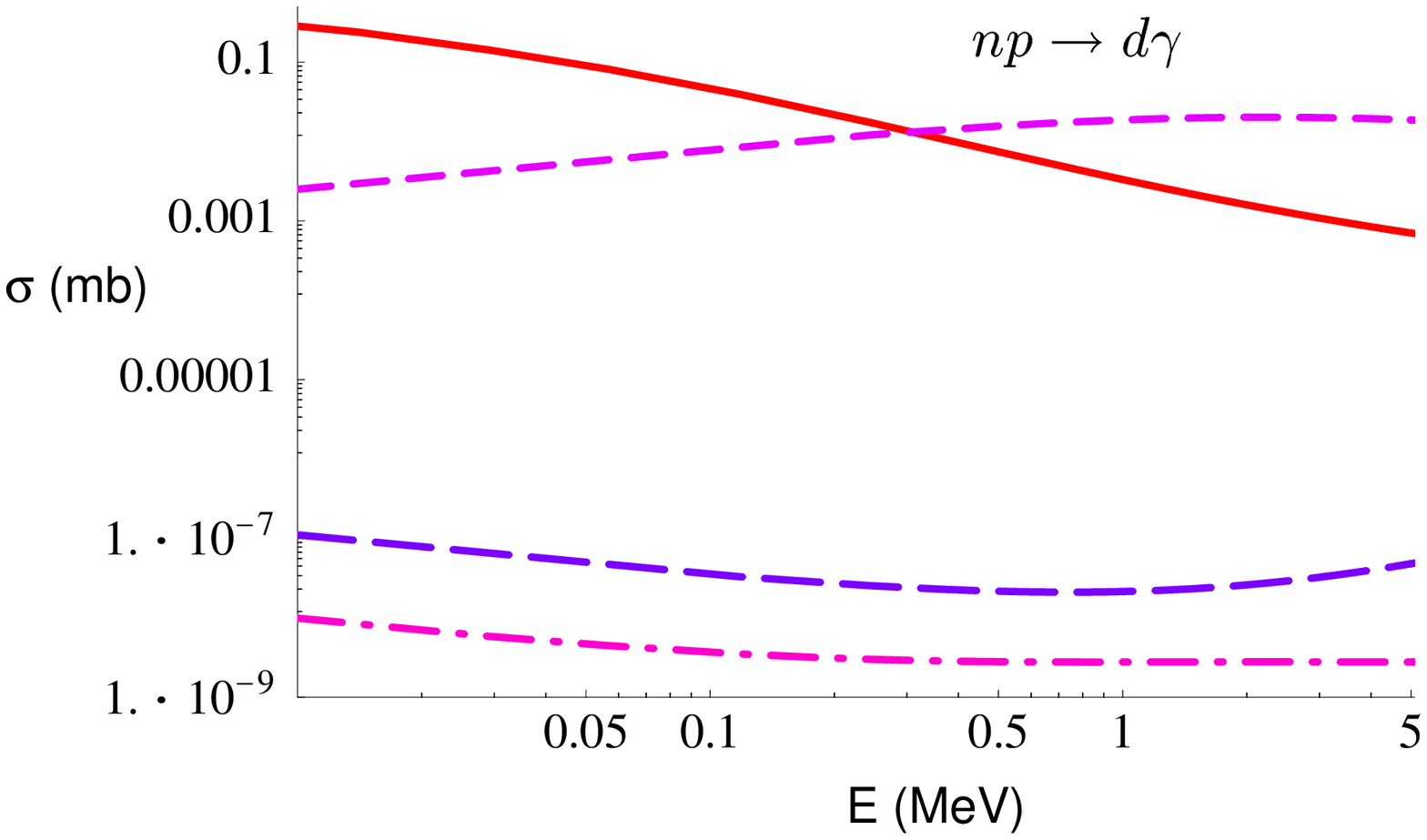}{2.5}{\protect $np\rightarrow d\gamma$ 
cross section $\sigma$ (mb) for center of mass energy E (MeV), on
a log-log plot. 
Solid curve: LO $M1_V$ transitions, dash: LO $E1_V$ transitions, long-dash: LO
 $M1_S$ transitions and dot-dash: LO $E2_S$ transitions.}

\begin{subsection}{Determining the Parameters $L_{np}$, ${\tilde L}_{np}$ and 
$L_{E1}$ }
\label{subsection:fit}
 The leading contributions
 from each partial wave are shown in Fig.~\ref{compare.fig}. One can see that 
the $M1_S$ and $E2_S$ transitions can be ignored for a $1\%$ level 
calculation in 
the energy range of interest.
At center of mass momentum $p_0=0.003443$ MeV the cross section for  
$np\rightarrow d\gamma$ is measured to be 
$\sigma^{exp}=334.2\pm 0.5$ mb~\cite{CWC} and $M1_V$ transitions give the 
dominant contribution, Fig.~\ref{compare.fig}. Thus at NLO, we fix $L_{np}$ 
from this cold neutron capture rate:
\beq\eqn{fitLnp}
\sigma^{(LO)}&\equiv&\frac{4\pi\alpha(\gamma^2+p^2)^3}{\gamma^3 M_N^4 p}
|{\tilde X}_{M1_V}^{(0)}|^2\ \Big|_{p=p_0}\nonumber\\ 
\sigma^{(NLO)}&\equiv&\sigma^{exp}-\sigma^{(LO)}=
\frac{4\pi\alpha(\gamma^2+p^2)^3}{\gamma^3 M_N^4 p}\ 2 Re({\tilde
X}_{M1_V}^{(0)}({\tilde X}_{M1_V}^{(1)})^\ast)\Big|_{p=p_0}\\ 
&&\Rightarrow L_{np}=-9.039\pm 0.027\ {\rm fm}^{2}\nonumber\ \cdot
\eeq
The N$^2$LO terms also contribute to the cross section at momentum $p_0$.  
Therefore, we impose the constraint~\cite{RS} 
\beq
\sigma^{(NNLO)}&\equiv&\frac{4\pi\alpha(\gamma^2+p^2)^3}{\gamma^3 M_N^4 p}\
\[|{\tilde X}_{M1_V}^{(1)}|^2+2Re({\tilde X}_{M1_V}^{(0)}({\tilde
X}_{M1_V}^{(2)})^\ast)\]\Big|_{p=p_0}=0\ , 
\eeq
to reproduce the experimental cross section $\sigma^{exp}$. This determines
 the parameter
\beq\eqn{fitLnptilda}
{\tilde L}_{np}=4.957\pm 0.011\ {\rm fm}^{2}\ \cdot
\eeq
The parameter
$L_{E1}$, which contributes only to $E1_V$ transitions, cannot be determined 
reliably from this single data point since the $E1_V$ cross section is 
negligible at this low momentum $p_0$.  
The error due 
to the experimental uncertainty in $\sigma^{exp}$ is significant for energies 
$E\lsim 0.25$ MeV. Above these energies the $E1_V$ cross section gives the
dominant contribution and the error in $M1_V$ cross section can be ignored.  
The contribution of the four-nucleon-one-photon
 operator is found to be significant at low momentum $p_0=0.003443$ MeV.
 Neglecting this operator by setting $L^{M1_V}_1=0$ at NLO underpredicts the
experimental result $\sigma^{exp}$ by $5\%$. This number is different from
the one quoted in Ref.~\cite{CRS1,SSW}, where the $\rho_d\gamma$ expansion was
used.

\makefig{breakup.fig}{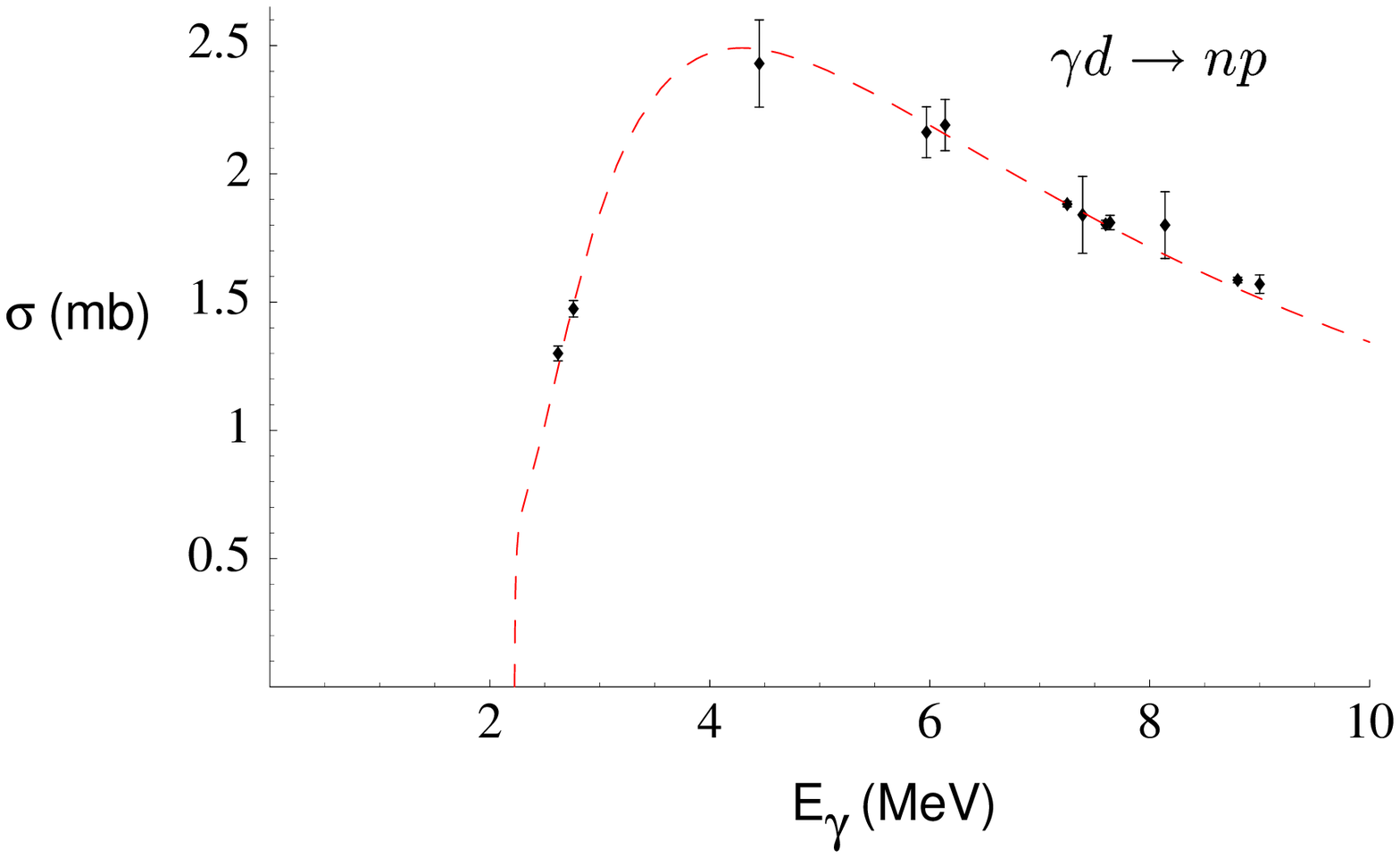}{2.5}{\protect  Cross section in mb for 
$\gamma d\rightarrow np$ as a function of photon energy $E_\gamma$ in MeV. 
The data is taken from Ref.~\protect\cite{book:AS}, pages 78 and 79.}

The parameter $L_{E1}$ can be determined from photodisintegration of the
deuteron. 
 This cross section is related to the neutron capture cross 
section by~\cite{CS,book:AS}:
\beq\eqn{fix:LE1}
\sigma(\gamma d\rightarrow np)\Align\frac{2M_N }{3 E_\gamma^2}\[E_\gamma 
- B +\frac{B^2-6 B E_\gamma +4 E_\gamma^2}{4 M_N}\ \]\sigma(np\rightarrow
d\gamma)\ , 
\eeq
where $E_\gamma$ is the incident photon energy in the deuteron rest frame. Only
the $E1_V$ transitions, on the right hand side of Eq.~\refeq{fix:LE1}, receive
relativistic corrections. In principle it is possible to determine all the  
three parameters $L_{np}$, ${\tilde L}_{np}$ and $L_{E1}$ from the deuteron 
breakup data. However, such a determination leads to 
significant uncertainty in the $M1_V$ cross section. This is because 
there are only two data points in the
nucleon energy range sensitive to $M1_V$ transitions, $0.3$ MeV $\lsim E\approx
E_\gamma-B\lsim 0.5$ MeV.  We find that a $\chi^2$ fit
 to data~\cite{book:AS} in the photon energy range $2.6$ MeV 
$<E_\gamma<7.3$ MeV does not give a reliable constraint on the parameters
 $L_{np}$, ${\tilde L}_{np}$ and $L_{E1}$.
Constraining $L_{np}$ and ${\tilde L}_{np}$ from $\sigma^{exp}$ as described 
above determines them more accurately. The other parameter $L_{E1}$ is
determined from a   
$\chi^2$ fit to data~\cite{book:AS} in the photon energy range $2.6$ MeV 
$<E_\gamma<7.3$ MeV, where experimental errors in the total cross section from
the $M1_V$ transition are
negligible. This gives: 
\beq\eqn{fitLE1}
L_{E1}\Align -(5.3\pm 3.6)\ {\rm fm}^{3}\ \cdot
\eeq
The error due to experimental uncertainty is $\lsim 1\%$ over the range of
 $E\lsim 5$ MeV. This means that for nucleon energies relevant for BBN, 
$E\lsim 0.2$ MeV, the experimental uncertainty is significant compared to 
theoretical error,
see Fig.~\ref{error.fig}. A few more high precision
measurements in the incident  
photon energy range $2.5$ MeV $\lsim E_\gamma\lsim 5$ MeV would provide
important constraints on $L_{E1}$ and 
determine the \protect{$np\rightarrow d\gamma$} cross section more accurately
at energies relevant for big-bang nucleosynthesis.  
 
The contribution from N$^4$LO is found to be $\lsim 3\%$ for incident photon
energies $E_\gamma\lsim 8$ MeV and is a small correction to the 
N$^3$LO result~\cite{CS}. This is better than the naive theoretical estimate 
for energies $E_\gamma > 4$ MeV. This low energy theory, which is formally 
valid 
for energies $E_\gamma\lsim 8$ MeV, seems to reproduce data well above its 
range of validity, see Fig.~\ref{breakup.fig}. 

Before discussing the results, it is important to check 
the self-consistency of our 
perturbative EFT calculation. The calculations
were carried out under certain assumptions about the sizes of the
couplings. The fact that the fitted couplings, 
Eqs.~\refeq{fitLnp},~\refeq{fitLnptilda} and~\refeq{fitLE1},
 have sizes comparable to
the theoretical estimates, Eqs.~\refeq{estimateLnp} and~\refeq{estimateLE1}, 
suggests that the perturbative expansion is under control to the order
calculated. This allows us to make reliable error estimates.

\makefig{error.fig}{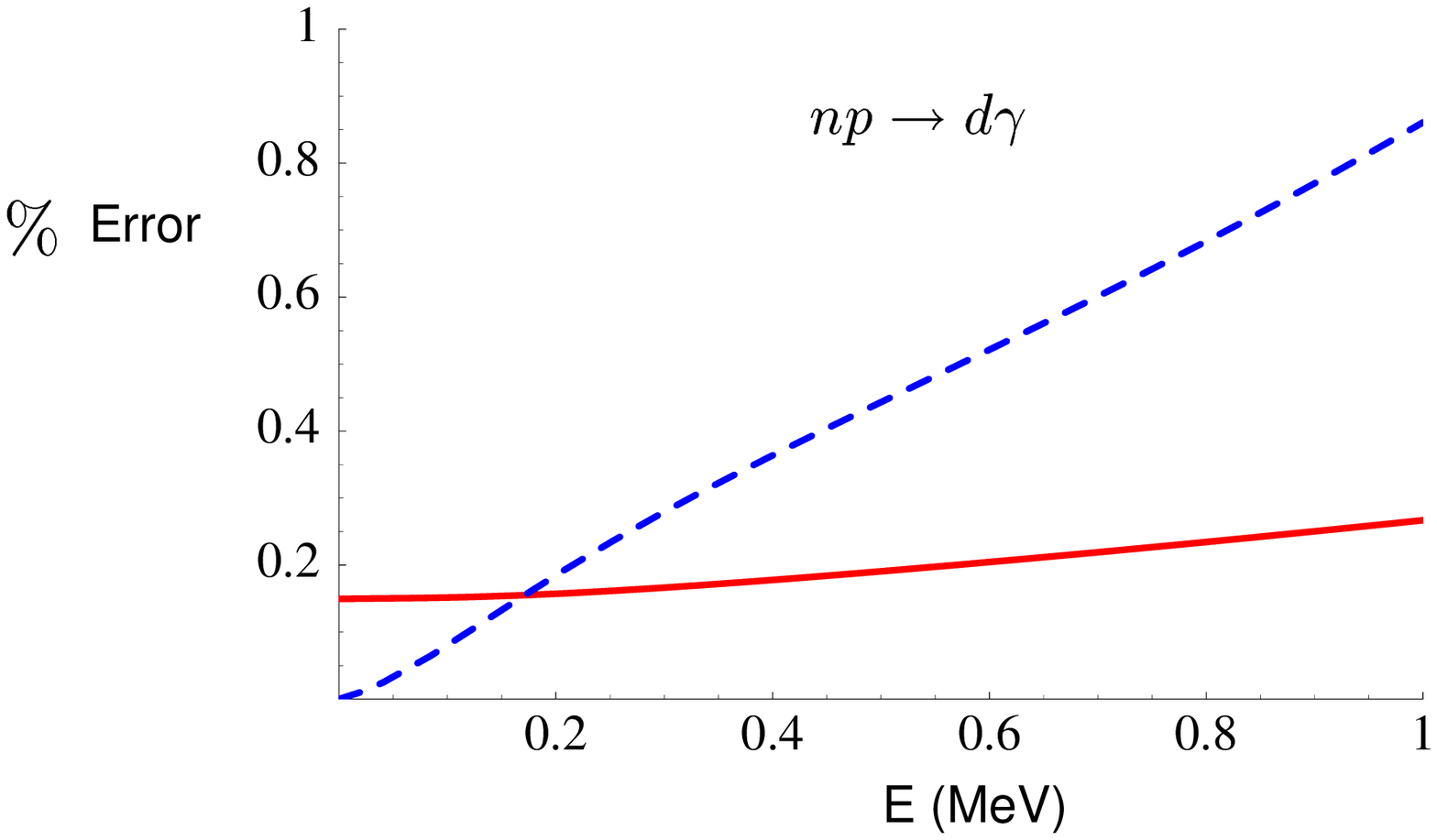}{2.4}{\protect  Estimated theoretical error
and experimental uncertainty in percentages versus nucleon center of mass
energy $E$ (MeV) for the $n p\rightarrow d\gamma$ cross section. Solid curve:
experimental uncertainty, dotted curve: estimated theoretical error.}   

\end{subsection}
\end{section}

\begin{section}{Results}
\label{section:results}
Table~\ref{table1} shows the EFT $np\rightarrow d\gamma$ cross section for 
various nucleon center of mass energies, $E$. The corresponding values for the 
cross section from the on-line ENDF/B-VI database~\cite{ENDF} are also shown in
the  
last column. As explained earlier, an error of 
$\(r_1 p^4/\gamma+r_0^3p^6/\gamma^3\)(1+|1/(a^{exp} p)|+|1/(a^{exp}\gamma)|)$
 with respect to LO was assigned to the $M1_V$ cross section and a
conservative error of  
$\gamma (p^4+p^2\gamma^2+\gamma^4)/(6\pi\Lambda^5) -
5(Z_d-1)\eta_d^2+(Z_d-1)(p^2+\gamma^2)/M_N^2$ 
 was assigned to the $E1_V$ cross section. 
The errors were added linearly to the total cross section and are found to be 
$\lsim 1\%$ for $E\lsim 1$ MeV. The EFT results are presented to only four 
significant digits, unless the theoretical errors enter earlier, in which case
we keep up to the first digit where the errors contribute. Compared to the
results 
in Ref.~\cite{CS}, these higher order EFT  results are
 perturbatively closer to the ENDF/B-VI values. However, at some energies, 
e.g. $1\times 10^{-3}$ MeV, where the difference between the N$^3$LO EFT   
result and ENDF value is much larger   
than the expected perturbative corrections, not surprisingly the 
discrepancy does not disappear by going to N$^4$LO~\cite{CS}.    
\begin{table}[htb]
\begin{tabular}{lllll}\multicolumn{5}{c}{ $\sigma(np\rightarrow d\gamma)$}\\ 
\hline
$E$ (MeV) & $M1_V$ (mb) & $E1_V$ (mb) & $M1_V+E1_V$ (mb)&
 ENDF (mb)~\cite{ENDF}\\ \hline
$1.2625\times 10^{-8}$ & $334.2\ ^*$ &
$5.122\times 10^{-6}$ & $334.2\ ^*$ & $332.0$ \\

$5.\times 10^{-4}$ & $1.667(0)$ & $0.001019(3)$ &
$1.668(0)$ & $1.660$ \\

$1.\times 10^{-3}$ & $1.170(0)$ & $0.001441(5)$ &
$1.172(0)$ & $1.193$ \\

$5.\times 10^{-3}$ & $0.4950(0)$ & $0.00322(1)$ &
$0.4982(0)$ & $0.496$ \\

$1.\times 10^{-2}$ & $0.3279(0)$ & $0.00454(2)$ &
$0.3324(0)$ & $0.324$ \\

$5.\times 10^{-2}$ & $0.09810(0)$ & $0.00997(3)$ &
$0.1081(0)$ & $0.108$ \\

$0.100$ & $0.04973(0)$ & $0.01379(5)$ &
$0.06352(5)$ & $0.0633$\\

$0.500$ & $0.00787(3)$ & $0.0263(1)$ &
$0.0341(2)$ & $0.0345$ \\

$1.00$ & $0.0036(1)$ & $0.0313(2)$ &
$0.0349(3)$ & $0.0342$\\
\end{tabular}
\caption{\protect Cross section for $np\rightarrow d\gamma$ in mb for 
different center of mass energy $E$ (MeV). For comparison, the values from the
ENDF/B-VI on-line database are also shown. The first entry ($^*$) is used for 
fitting a combination of parameters $L_{np}$'s, 
Eq.~\refeq{fitLnp},~\refeq{fitLnptilda}. Another 
parameter $L_{E1}$ was fitted to the deuteron 
photodisintegration
 $\gamma d\rightarrow np$ cross section, Eq.~\refeq{fitLE1}.  \label{table1}}
\end{table}

\end{section}

\begin{section}{Summary and Conclusions}
\label{section:conclusions}
In finale, the big-bang nucleosynthesis prediction for primordial light
element abundances uses the cross section for $np\rightarrow d\gamma$ as 
an input. A  
precise estimate of this cross section is thus critical for these predictions. 
The total cross section for 
radiative capture of neutrons on the proton $np\rightarrow d\gamma$ was
 calculated
 in EFT for center of mass energies $E\lsim 1$ MeV. At the energies relevant 
for BBN, $0.02$ MeV $\lsim E\lsim 0.2$ MeV,
the isovector electric transition $E1_V$ and the isovector magnetic transition
$M1_V$ give the dominant contributions. 

We calculated the $E1_V$ cross section to N$^4$LO and the $M1_V$ cross
 section to N$^2$LO. 
 Up to N$^3$LO, the
 $E1_V$ amplitude calculated in EFT is equivalent to the effective range 
theory result. Thus in principle, any calculation including potential models
that 
reproduces nucleon-nucleon scattering data well will be able to describe the 
$E1_V$ amplitude to the accuracy of the N$^3$LO calculation.  
However, the effective range theory result differs from the 
N$^4$LO EFT result due to the absence of a four-nucleon-one-photon 
operator. This operator cannot be obtained from nucleon-nucleon scattering data
alone. 
We determine this operator from the deuteron breakup
$\gamma d\rightarrow np$ data with significant experimental uncertainty.
Similarly, for the $M1_V$  
transition there is a four-nucleon-one-photon operator at NLO that is not 
related to nucleon-nucleon scattering operators by gauge transformations.
We determine this unknown operator from the cold neutron capture 
$np\rightarrow d\gamma$ rate and
this operator contributes $\sim 2\%$ to the total cross section at $E=0.5$ MeV.
A few more precise measurements in the incident
photon energy range $2.5$ MeV $\lsim E_\gamma \lsim 5$ MeV would provide
important constraints on the $M1_V$ and $E1_V$ transitions in the energies
relevant for BBN. The sizes of the operators determined from data are
consistent 
with their theoretical estimates. This verifies the theoretical assumptions
about  
the perturbative expansion to the order of the calculation, which allows us to
make reliable error estimates.   
For energies $E\lsim 1$ MeV, the theoretical uncertainty in the total cross
section is estimated to be $\lsim 1\%$.

\end{section}

\begin{section}*{acknowledgment}
The author would like to thank Jiunn-Wei Chen, Daniel Phillips and Martin 
Savage for many helpful discussions. I would also like to thank Paulo Bedaque
 and Noam Shoresh for useful comments on the manuscript, and the other
members of the EFT 
group at the University of Washington for valuable suggestions.   
 This work is supported in part by
 the U.S. Dept. of Energy under Grant No. DE-FG0397ER4014.
\end{section}


\end{document}